# Design of on-chip plasmonic modulator with vanadium-dioxide in hybrid orthogonal junctions on Silicon-on-Insulator

*Gregory Tanyi[1*], Miao Sun[1*], Ranjith Unnithan[1*]*

[1]Department of Electrical and Electronic Engineering, University of Melbourne, Parkville, Victoria 3010, Australia
[*]E-mail: gtanyi@student.unimelb.edu.au, aussunnysun@gmail.com, r.ranjith@unimelb.edu.au

We present a plasmonic modulator based on hybrid orthogonal silver junctions using vanadium dioxide as the modulating material on the silicon on insulator. The modulator has an ultra-compact footprint of 1.8μm x 1μm with a 100nm x 100nm modulating section based on the orthogonal geometry. We take advantage of large change in the refractive index of vanadium dioxide during its phase transition to achieve a high modulation depth of 46.89dB/μm. We also provide a fabrication strategy for the development of this device. The device geometry has potential applications in the development of next generation high frequency photonic modulators for optical communications which require a nanometer scale footprint, large modulation depth and small insertion losses.

## 1. INTRODUCTION

Advances in nanofabrication in the past 30 years have led to the development of more compact and faster photonic and electro-optic devices [1]- [3]. Modulators are electro-optic devices that encode a high-speed electronic data stream to an optical carrier wave in photonic integrated circuits for optical communications. The size, power consumption and the frequency of operation are the key metrices used to evaluate performances of the modulators. Recently, there has been a surge in the development of novel modulators based on silicon photonics because both CMOS and the silicon photonics are based on silicon. This compatibility enables the integration of photonics circuits and electronic circuits in a single chip to increase the speed of operation and to reduce footprint.

However, reducing the footprint while keeping the speed of operation is still a challenge. This is because the footprints of conventional silicon photonics optical devices are limited by the diffraction limit. Furthermore, silicon does not exhibit a linear electro-optic (EO) effect and hence the silicon modulators are either based on a dispersion effect which alters the carrier concentration of Si to tune its permittivity or by integrating it with materials which display an EO effect. This requires a large interaction length (footprint) between the electrical signals and the optical signals. To circumvent these constraints, plasmonic- based devices (surface plasmon polaritons (SPPs)) have been integrated in silicon photonics. SPPs are electromagnetic surface waves at a dielectric–metal interface, coupled to the charge density oscillation in the metal surface [23]. SPPs offer the ability to focus light on nanoscales and are key elements in the development of subwavelength optical components with the added advantage of being compact and operating at much high frequencies [20]. Recently, Graphene based plasmonic modulators have been used to achieve modulation efficiency of 0.417dB/μm [17]. Also, plasmonic modulators based on the Pockels effect and operating at 40GHz have been developed using electro-optical polymers with a footprint of 29μm [16]. Designing plasmonic devices usually entails a trade-off between modulation depth, device size, loss and extinction ratio [16]. Using EO materials without a large enough refractive index change require more interaction length between modulating radio frequency (R.F) signal and optical laser signal in the device, which lead to large footprint, high radiative and dissipative losses [16]. Vanadium dioxide is a canonical mott material which exhibits a first order insulator to metal transition (IMT) which can be triggered by exciting the material thermally, electrically, optically and by doping [20]. This phase transition is accompanied by a corresponding large change in refractive index in both phases across the bulk material.

It has been experimentally shown that an electric field strength of $6.5\times10^7$ V/m would trigger the insulator to metal transition of $VO_2$ [19]. This phase transition occurs on the femto second scale (26fs) and as such, $VO_2$ has been used as the modulating material in many plasmonic devices. In [26], Thermally driven switches based on $VO_2$ have been developed with extinction ratios of 6.4dB/µm and a 5µm active region. However, thermally driven VO2 devices have a limited speed of operation. $VO_2$ based modulators have been explored in [7],[9],[19]. However, some of these devices have a large footprint whereas others require an initial heat to trigger the change in phase of $VO_2$.

In this paper, we report a hybrid orthogonal plasmonic modulator with a 100nm x 100 nm modulation section within a compact device footprint of 1.8µm x 1.0 µm. The electro-optic material of choice is vanadium dioxide. There is a large change in the refractive index of $VO_2$ with the semiconductor phase having a refractive index of 3.24+0.30i and the metallic phase having a refractive index of (2.03+2.64i). The optical modulation in this device is achieved by the large refractive index change of the nanoscale $VO_2$ in the plasmonic slot due its phase transition. The orthogonal coupling geometry makes the device footprint small and provides a high modulation index of 46.89dB/µm at telecommunication wavelength, 1550 nm. Furthermore, the orthogonal geometry parameters are investigated and optimized for the modulator development. We also propose the fabrication strategy to show that the proposed geometry can be fabricated using the existing micro and nanofabrication capabilities.

## 2. General Device Description

The proposed modulator geometry is shown in the Fig.1. below. The coupling scheme used is similar to the published work in [2].

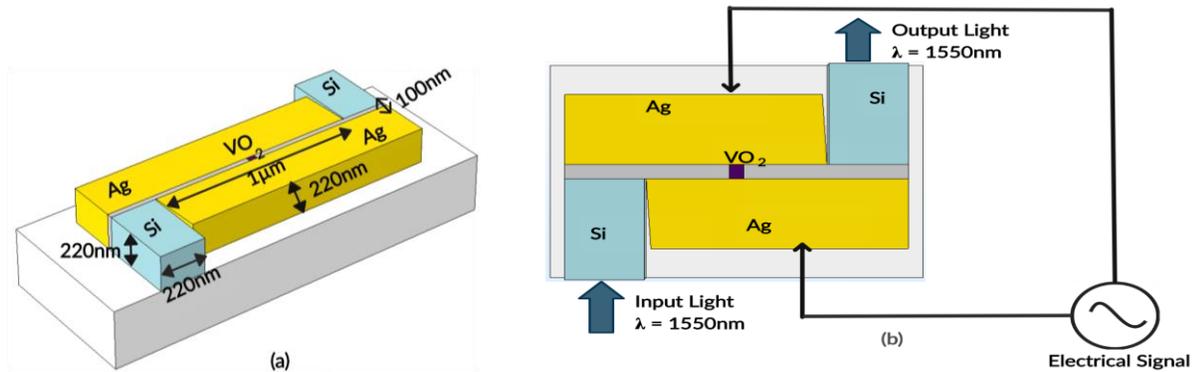

*Fig. 1.(a) Three-dimensional geometry of the proposed hybrid orthogonal plasmonic modulator. (b) Two-dimensional geometry of the proposed hybrid orthogonal plasmonic modulator. (Top layer of air not shown for visibility)*

In this geometry, the light of wavelength 1550nm travels in a silicon waveguide (width 430nm and height 220nm) and is then coupled at the orthogonal silicon-air junction to a plasmonic slot waveguide (100nm width and 1µm length). The plasmonic slot is made of 220nm thick silver due to its relatively low plasmonic losses [23]. A 100nm long section of $VO_2$ is introduced at the center of the plasmonic slot as the modulating section. The silver electrodes are extended via wire bonding and connected to an external voltage source to apply voltage across the slot to change the phase of $VO_2$. At the second orthogonal junction, the plasmons are coupled back into photons and travel along the output silicon waveguide. A grating coupling scheme (not shown in the figure above) is used for coupling in light from an optical fibre to the silicon waveguide for coupling the light out into another optical fibre.

The device operates in 2 states depending on the application of an external electric field which drives the insulator to metal transition and vice versa of the $VO_2$ in the modulating section. In the OFF state, there is no external electric field and hence $VO_2$ in the modulating section is in the semiconductor phase with a refractive index of (3.24+0.30i). In this phase, surface plasmons polaritons travel across the modulating section and interact with $VO_2$ in its semiconductor phase. In the ON state, there is a high enough electric field to trigger the insulator to metal transition (IMT) of $VO_2$. In the metallic phase, $VO_2$ has a lower refractive index but higher extinction coefficient (2.03+2.64i) and which leads to the surface plasmons polaritons being attenuated by the $VO_2$ section. The surface plasmons propagating in the slot through $VO_2$ get modulated as it toggles between the ON and OFF states. The modulation depth is described as how much the modulation variable of a propagating carrier varies around it's normal unmodulated level. Here, the modulation variable is the optical loss (attenuation). The optical loss is measured as the ratio of output power to the input power which are obtained by calculating the surface integral of the optical power at the silicon waveguides (input port and output ports).

In the orthogonal modulator, the silicon waveguide has a width of 430nm and a height of 220nm, the angular separation (theta) between the silicon waveguide and the silver electrode is 10°, the width of the plasmonic slot (w) is 100nm and the length of $VO_2$ (L) used is 100nm. Silver electrodes of height 220nm are used as walls of the plasmonic slot and the middle of the slot is filled with a 100nm long section of $VO_2$. The above parameters are optimized for the C-band frequency range of operation of the device as shown in Fig.1 (b) (OFF state of the device is shown). The simulation details used to obtain the optimal parameters for the modulator along with the results are discussed the in the section below.

### 3. Modelling and simulation results

We perform the modulator design, simulation and optimization using the finite element methods implemented in the COMSOL Multiphysics commercial software. A minimum mesh element size of 6nm is used in the plasmonic slot section which contains the smallest device features. Scattering boundary conditions (SBCs) are utilized to terminate the computational domain. The refractive indices of $VO_2$ were obtained with the aid of variable angle spectroscopic ellipsometry as presented in [24].

Fig. 2(a) shows a top view of the modulator in the OFF state. In this state, light in the silicon waveguide is coupled to the plasmonic slot and then propagates as SSP through the slot interacting with $VO_2$ in the semiconductor phase before being coupled back into the light at the output waveguide. Fig. 2(b) shows the same top view in the ON state. During the ON state, the metallic $VO_2$ attenuates the surface plasmons significantly by reducing the intensity of light coupled back to the output waveguide.

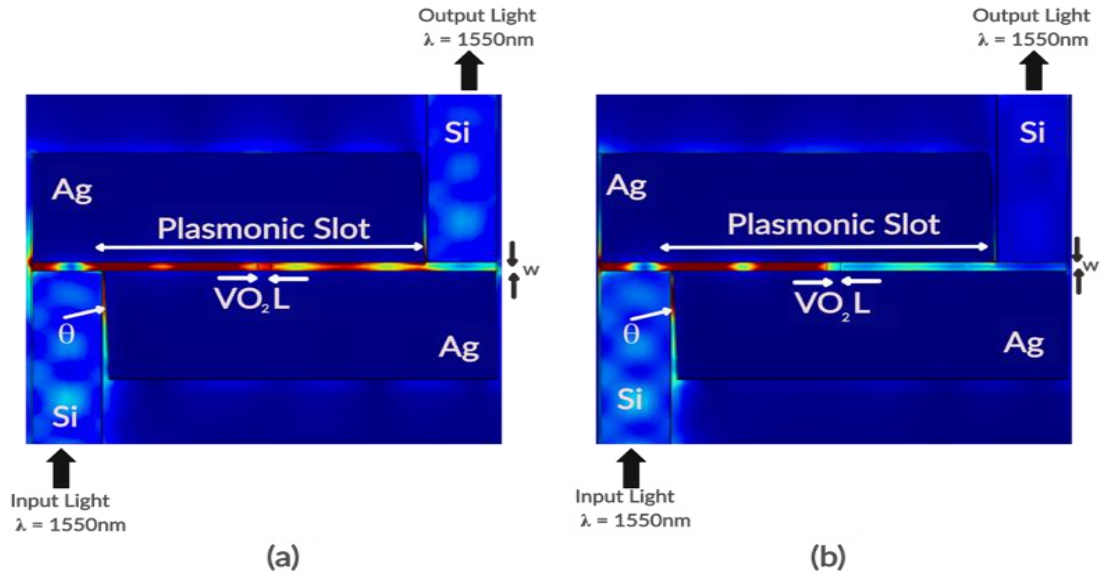

*Fig. 2 Planar cross-section view of device showing surface plot of electric field norm in (a-Left) Semiconductor phase of $VO_2$ (Device OFF) and (b-Right) Metallic phase (Device ON) of $VO_2$.*

Figures 3(b) and 3(c) show a cross section of the electric field along the Ag-$VO_2$-Ag plasmonic slot with a strong confinement of the electric field within the slot. It is observed that in the ON state of the device, the electric field intensity is lower than in the OFF state because of the higher optical loss which accompanies the phase change of $VO_2$ to the metallic phase. We further examine the E field confinement in both states by examining the electric field intensity along a cutline in the modulating section of the plasmonic slot. The electric field intensity profiles are extracted from simulations with the cutline taken in the geometric perpendicularly to the plasmonic slot as shown in figures 3(b) and 3(c). Figure 3(a) shows that in the ON state, there is a significant drop in electric field intensity which is consistent with figure 2 given the high extinction coefficient of $VO_2$ in its metallic phase. As such, Fig3 provides a detailed view of the changes in the modulating section of the device with a change from the OFF state to the ON state.

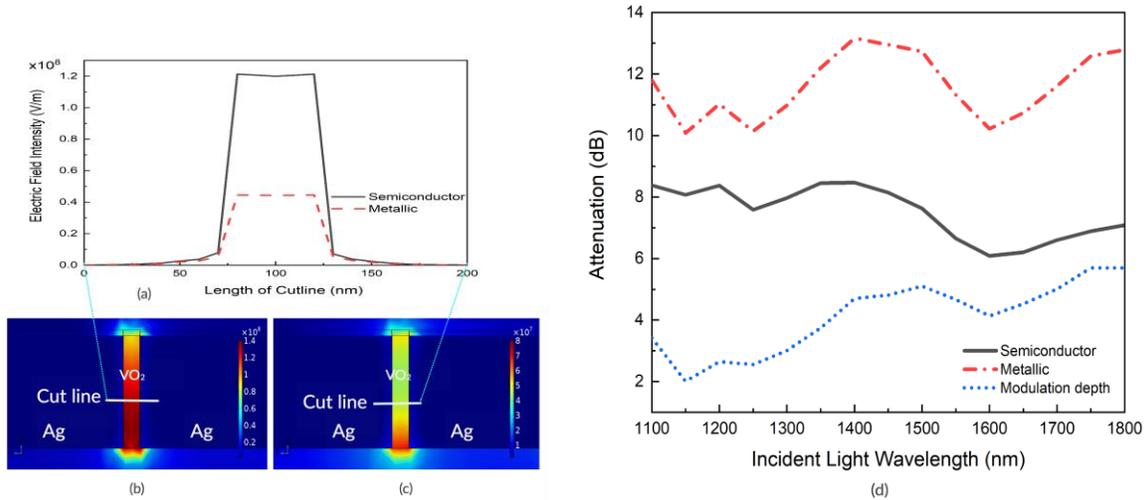

*Fig. 3(a)Electric field Norm variation along cutline through plasmonic slot showing strong field confinement, Longitudinal cross section of the device in (b) OFF state ($VO_2$ in semiconductor phase) (c) ON state ($VO_2$ in metallic phase) (d) Insertion loss in both the metallic phase (ON phase) and the semiconductor phase (OFF phase) for incident light wavelengths ($\lambda$) from 1100nm to 1800nm.*

Figure 3 (d) shows the optical loss of the device in both states along the wavelengths (1100nm – 1800nm). The device is thus optimized to have a minimal insertion loss around the C Band of the telecommunications window.

A key parameter that affects the level of modulation of our device is the length (L) of $VO_2$ used. In our simulations, we choose to perform a parametric sweep of L values from 50nm to 170nm at the 1550nm wavelength because within this range the fabrication is feasible, and the optical loss is tolerable. Fig. 4(a) shows how the attenuation and modulation depth vary with the length of $VO_2$ at the 1550nm wavelength which is our wavelength of interest because of its telecommunication applications. We observe a linear increase in attenuation as the length of $VO_2$ in the slot is increased. This can be explained by the extinction coefficient of $VO_2$ in both phases. Given $VO_2$ is not a transparent material, the plasmons are attenuated when they interact with the modulating section and as such, increasing the length of $VO_2$ means more loss in the plasmon energy. The higher loss in the metallic phase is because of a higher extinction coefficient in the metallic phase. We also study the effect of varying the wavelength on the device performance. For this, we perform a wavelength sweep from 1100nm to 1800nm and compare the attenuation in both the metallic and semiconductor phases. The results as shown in Fig.4(b) show the device is robust enough to handle wavelength shifts because of transient temperatures in high power laser systems. The trend of increased attenuation with an increase in length of $VO_2$ is consistent in both figures Fig.4(a) and Fig.4(b). The slight dips at the 1150nm and 1600nm wavelengths are from the refractive index data used for the simulation obtained from literature [24].

To obtain the attenuation, we first calculate the ratio of the surface integral of the optical power Poynting vector (S) at both ends of the silicon waveguide which are the input and output ports [24]. The Poynting vector is obtained as $S[W/m^2] = \hat{n}I$ [20]. The surface integral of the optical Poynting vector at the input and output ports give us the input power ($P_{input}$) and output power ($P_{output}$) respectively. The optical power is given by $P = \iint \vec{S} \cdot \hat{n}\, da$ [20]. The attenuation in dB is given by: $Attenuation = -10 \log_{10}\left(\frac{P_{output}}{P_{input}}\right)$ [20]. The modulation depth (dB) is given the difference in attenuation in the OFF state and the ON state.

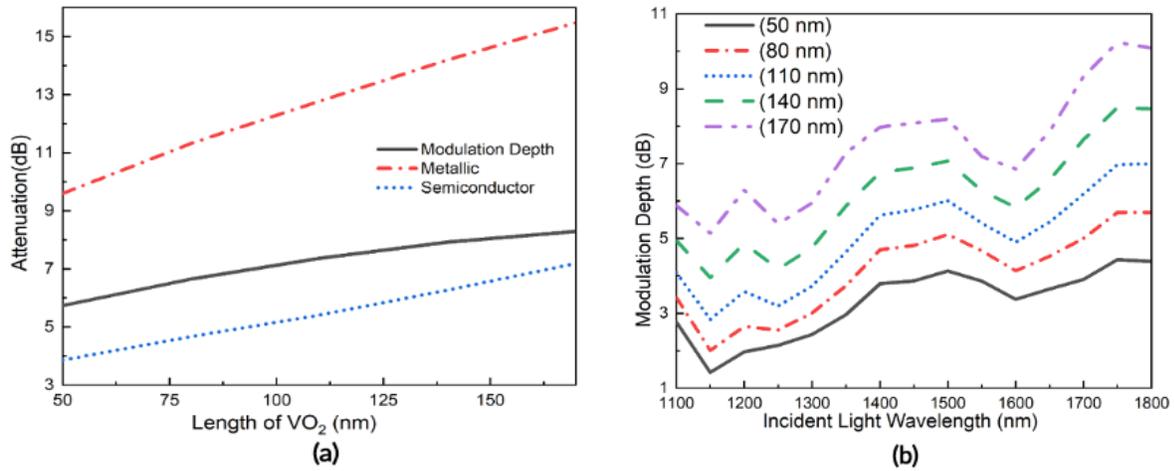

*Fig. 4 (a) Attenuation of light of wavelength 1550nm for both the metallic and semiconductor phases of VO₂ as well as the modulation depth for different lengths of the VO₂ modulating section (L). (b) The variation of the modulation depth with different wavelengths of incident light (λ) from 1100nm to 1800nm, with the length of the VO₂ modulating section (L) varied from 50nm to 170nm.*

In the real fabrication of the device, there is a need to understand the impact of the width of the plasmonic slot on the device performance as well as the angle separating the electrodes from the waveguides as these could slightly vary because of limitations of the current fabrication technology. We thus study the impact of increasing the width of the plasmonic slot (w) on the attenuation in both phases and the modulation depth. Fig 5(a) shows there is a decrease in modulation depth as the width is increased. This is because increasing the width of the plasmonic slot reduces the photon to plasmon coupling efficiency.

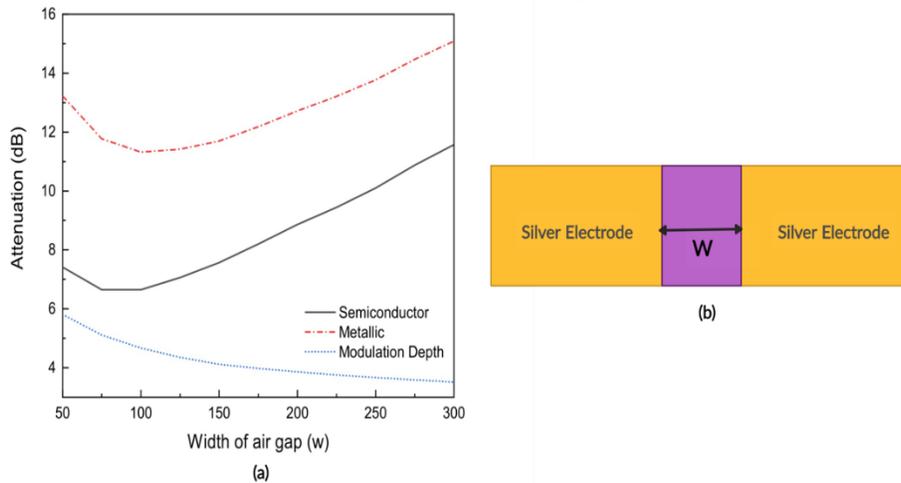

*Fig. 5 (a) Attenuation of Light at 1550nm for the metallic and semiconductor phases of VO₂ with the modulation depth for different widths w of the plasmonic slot. (b) Cross section of the plasmonic slot of width w with violet colour representing VO₂ and yellow colour representing Ag electrode.*

The dependence of the modulation depth of the device on the angular separation is studied by sweeping the angular separation (θ) from 10° – 85 °. From our simulation results, we observe that the modulation depth reduces as the angular separation is increased. This is because a smaller angular separation accounts for a higher coupling efficiency.

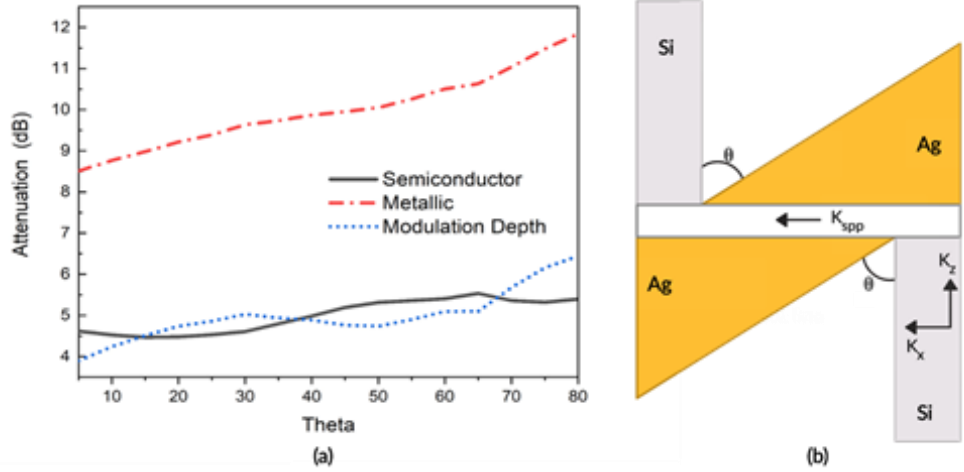

*Fig. 6 Attenuation of Light at 1550nm for the metallic and semiconductor phases of VO2 with the modulation depth for different angular separations (theta) between the Ag electrode and the silicon nanowire. (b) 2D-schematics of device with Silver highlighted in yellow and silicon highlighted in grey along with transverse component of silicon nanowires momentum ($k_x$), orthogonal component of silicon waveguide momentum($k_y$) and surface plasmon polariton momentum annotated ($k_{spp}$).*

Using only a 100nm modulation section of $VO_2$ and with the following optimal parameters (w= 100nm, θ=8º), we achieve a modulation depth of 4.69 dB (46.89dB/μm). From figures 2 and 3 above, the high extinction coefficient plays a key role in achieving this high modulation depth.

We further investigate the effect of the phase change of $VO_2$ on the coupling efficiency of this hybrid orthogonal geometry. In this hybrid-orthogonal geometry, the coupling efficiency is highly related to the momentum mismatch between the SPP mode and the waveguide [7]. Maximum coupling occurs when the orthogonal component of the momentum of the waveguide ($k_x$) matches the fundamental mode of the plasmonic slot ($K_{spp}$) and there is thus a minimal spatial mismatch. The schematics of this coupling scheme are illustrated in Fig 6(b). In figure 7, the longitudinal wave vector (and thus momentum) of the plasmonic slot with $VO_2$ in both the metallic phase and semiconductor phase is compared to the orthogonal component of the waveguide's wave vector (and thus momentum). Figure 7 shows that at the 1550nm wavelength, the plasmonic slot with $VO_2$ in the semiconductor phase $K_{spp}$ equals $K_x$. Also, observing both curves shows the device has a higher coupling efficiency with $VO_2$ in the semiconductor phase than the metallic phase as the device was optimized for the semiconductor phase. This change in coupling efficiency as well as the high extinction ratio of $VO_2$ lead to the high modulation depth in the results above.

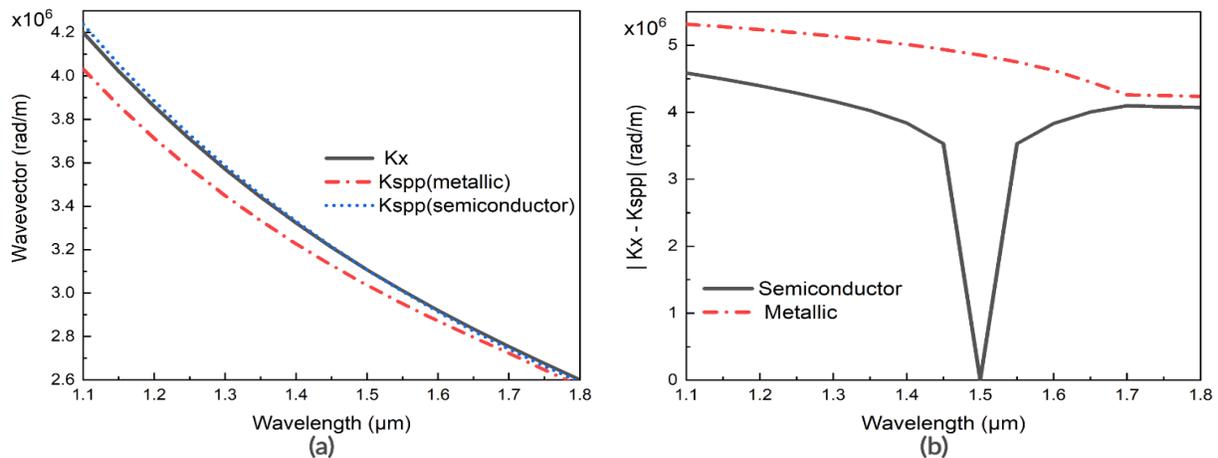

*Fig. 7 (a) Dispersion characteristics of the plasmonic slot and the silicon waveguide with VO2 in both the metallic and semiconductor phases. The slot width (w) is 100nm, the silicon waveguide is 430nm wide and 220nm high (b) Absolute value of spatial mismatch in the metallic (ON) and metallic (OFF) states showing optimal coupling in the OFF state at the 1550nm window.*

Fabrication strategy

Our proposed design can be fabricated using the standard Silicon on Insulator (SOI) wafer with a 220nm thick silicon layer as the top layer sitting on a 2μm layer of silicon dioxide sitting on a handling layer of silicon (550μm). The fabrication methodology can be divided in 4 steps. Firstly, the silicon waveguides are patterned using electron beam lithography [25] with a double layer of HSQ resist (440nm). The patterned waveguide is then developed with the salty developer and etched using DRIE etching with the $SiO_2$ from the developed HSQ used as a mask. The second step involves the fabrication of the silver electrodes. We propose spin coating the substrate with PMMA A4 (440nm) before patterning the electrodes [26] using electron beam lithography. Electron beam evaporation is then used for the deposition of 220nm of Ag with Chromium used as an adhesive layer which is washed away during a liftoff process. The third step involves using the focus ion beam scanning electron microscopy to mill the plasmonic slot between the electrodes. The final step involves the physical vapor deposition of $VO_2$ into the plasmonic slot. The figure below summarizes the proposed fabrication strategy.

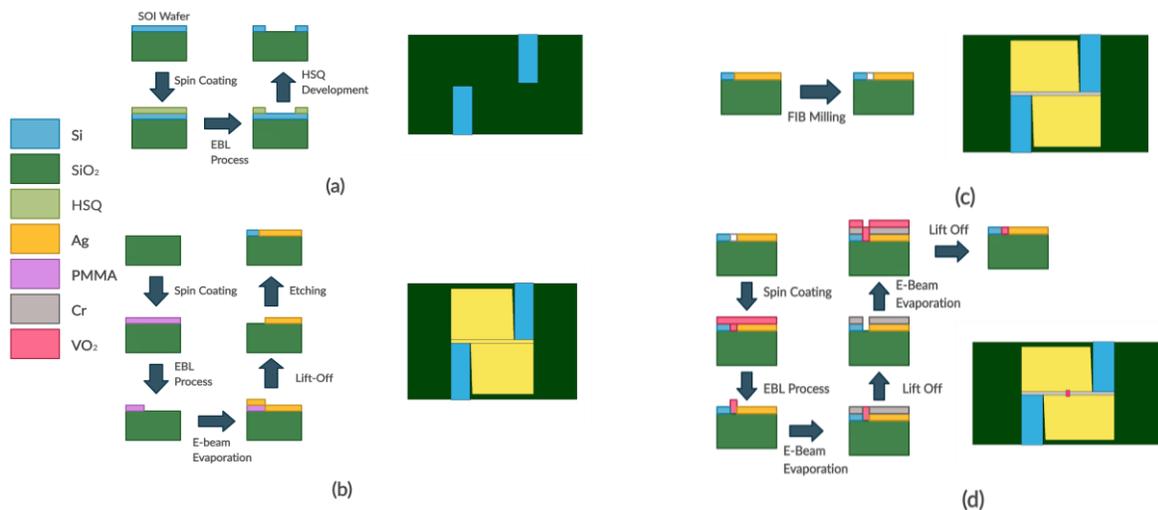

*Fig. 8 Layout of Fabrication Strategy for the device. (a) Silicon Waveguide (b) Silver electrodes. (c) Plasmonic slot milling and (d) $VO_2$ deposition.*

conclusion

In this paper, we have demonstrated a hybrid orthogonal plasmonic modulator with a compact 100nm x 100nm modulating section and a small device footprint of 1.8μm x 1μm. The device used Vanadium dioxide ($VO_2$) as the electrooptic material for intensity modulation and exploit the large change in the refractive index of $VO_2$ as it undergoes an insulator to metal transition. The modulator is optimized for the 1550nm telecommunications wavelength but can operate broadly in the O, E, S, C, L and U Bands (1260nm – 1675nm) and outperforms similar devices in literature to achieve a modulation depth of 46.89dB/μm at the 1550nm. We also study the coupling mismatch in the ON and OFF device states by exploring the momentum mismatch in the device as VO2 switches from the semiconductor to the metallic phase. This device can be readily fabricated, and we propose a fabrication strategy. The results will have applications in design of compact, cutting-edge high frequency modulators for high-speed optical communications.

**Data availability:**

The data that support the findings within this paper are available from the corresponding authors on reasonable request.